\documentclass[fleqn,12pt,twoside]{article}
\usepackage{espcrc1,epsfig}

\usepackage{graphicx}
\usepackage[figuresright]{rotating}

\newcommand{\AmS}{{\protect\the\textfont2
  A\kern-.1667em\lower.5ex\hbox{M}\kern-.125emS}}

\title{Perspectives in Galactic Chemical Evolution studies}

\author{N. Prantzos\address[IAP]{Institut d'Astrophysique de Paris, 
        98bis Bd Arago, 75014 Paris}}
       
\begin{document}

\maketitle

\begin{abstract}
In this review I focus on a few selected topics, where 
recent theoretical and/or observational progress has been made and important
developments are expected in the future. They include: 
1) Evolution of isotopic ratios, 2) Mixing processes and dispersion in 
abundance ratios,  3) Abundance gradients in the Galactic disk (and abundance
patterns in the inner Galaxy), 
4) The question of primary Nitrogen and 5) Abundance patterns in 
extragalactic damped Lyman-$\alpha$ systems (DLAs).
\end{abstract}

\section{INTRODUCTION}

Despite the lack of a reliable model of Galactic Chemical Evolution (GCE)
considerable progress has been made in the past few years, due to:

- an impressive amount of observational data concerning abundance ratios
either in the Milky Way (Carretta et al. 2002,
Truran et al. 2002, Andriefsky et al. 2002, and references therein) 
or in extragalactic systems (e.g. Prochaska and Wolfe 2002);

- the publication of detailed nucleosynthesis yields
(e.g. van den Hoek and Groenewegen 1997 for intermediate mass stars;
Woosley and Weaver 1995, Nomoto et al. 1997, Limongi et al. 2000, for 
massive stars; Iwamoto et al. 1999 for SNIa).

Interpreted in the framework of simple ``toy'' models of GCE these data offer
mainly information about: i) evolutionary timescales of corresponding 
production sites, ii) physics of those sites, iii) nature of relevant
nucleosynthetic processes.

In this, highly biased, review I focus on a few selected topics, where 
recent theoretical and/or obervational progress has been made and important
developments are expected in the future.

\begin{figure*}[htb]
\begin{center}
\psfig{figure=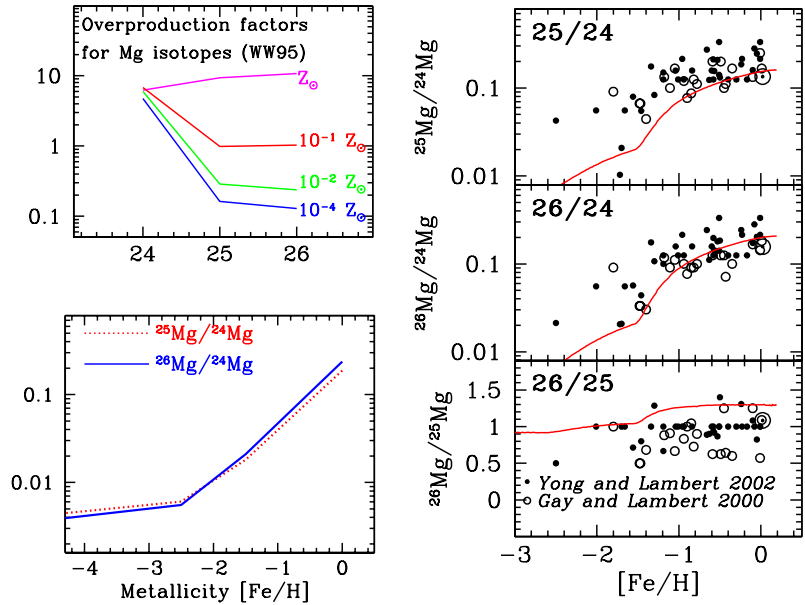,angle=-90,width=0.7\textwidth}
\caption{{\it Top left:} Overproduction factors of the Mg isotopes as
a function of stellar mass and for various initial metallicities (from WW95).
{\it Bottom left:} Abundance ratios of the Mg isotopes (integrated over
the Kroupa et al. 1993 IMF) as a function of initial stellar metallicity.
{\it Right:} Evolution of Mg isotopic ratios in Milky Ways' halo and
local disk stars (from Goswami and Prantzos 2000) 
with WW95 yields and comparison to observations.
}
\end{center}
\end{figure*}

\section{EVOLUTION OF Mg ISOTOPIC RATIOS IN THE MILKY WAY}

There are very few cases where observations allow to check models
of isotopic abundance evolution in the Galaxy, especially concerning
the early (i.e. halo) phase of that evolution. One of these rare cases
concerns the Mg isotopes, which are mainly produced by hydrostatic burning
in the carbon and neon shells of massive stars. The production of the
neutron-rich isotopes $^{25}$Mg and $^{26}$Mg is affected by the 
neutron excess (their yields increase with initial stellar metallicity)
while $^{24}$Mg is produced as a primary (its yield is
independent of metallicity).  Thus, the
isotopic ratios $^{25}$Mg/$^{24}$Mg and $^{26}$Mg/$^{24}$Mg are expected
to increase with metallicity (Fig. 1, left panel).

\begin{figure}[t]
\begin{minipage}[t]{75mm}
\psfig{figure=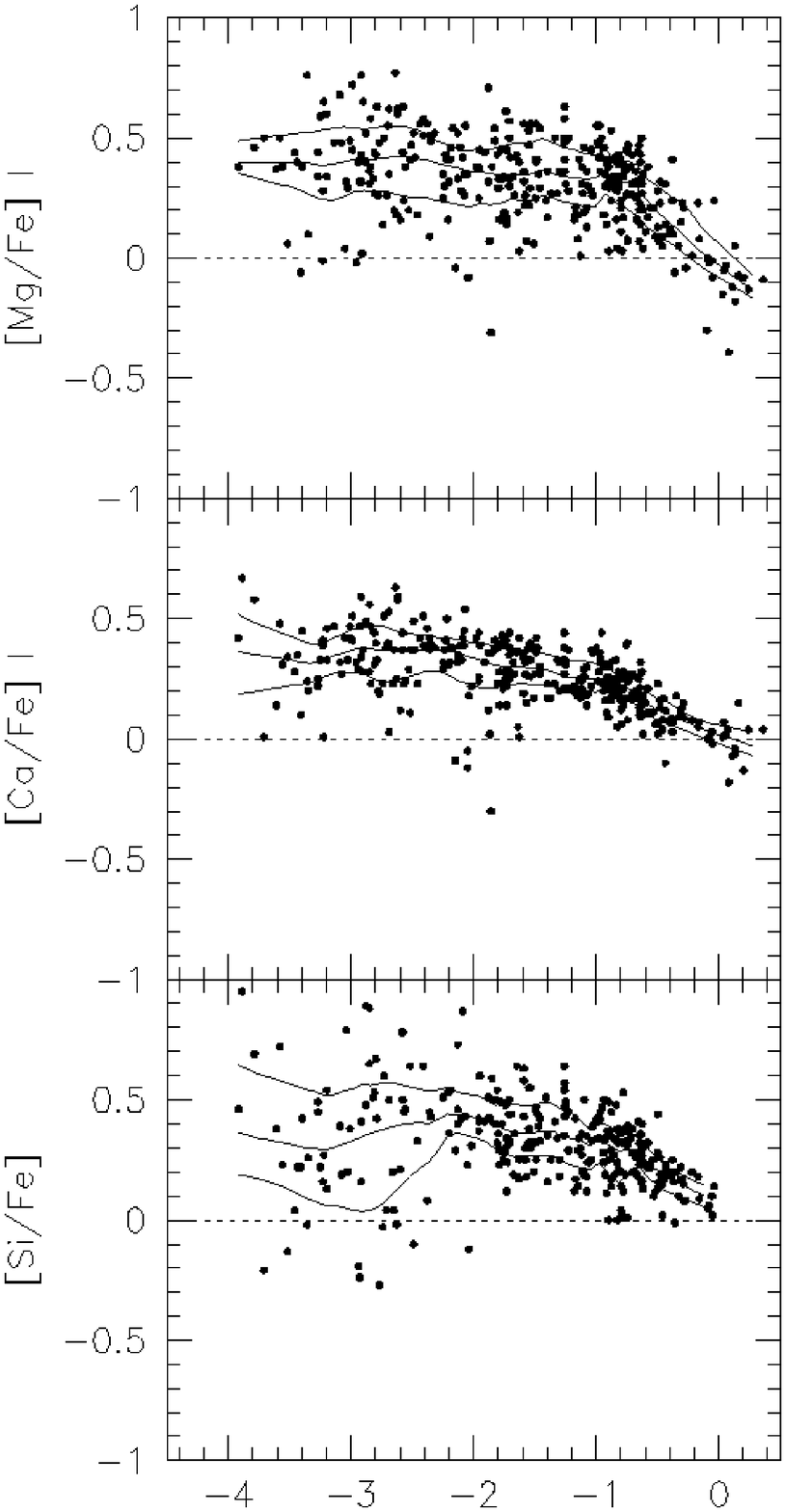,angle=0,height=7cm,width=\textwidth}
\caption{Abundance ratios [M/Fe] vs [Fe/H] for halo and local disk
stars; curves in the middle  indicate mean trends, while upper and lower
curves indicate ~1 $\sigma$ scatter (from Carretta et al. 2002). There is
no hint for intrsinsic dispersion in the data, if account is taken of the
observational uncertainties of $\sim$0.1 dex.} 
\label{fig:largenenough}
\end{minipage}
\hspace{\fill}
\begin{minipage}[t]{75mm}
\psfig{figure=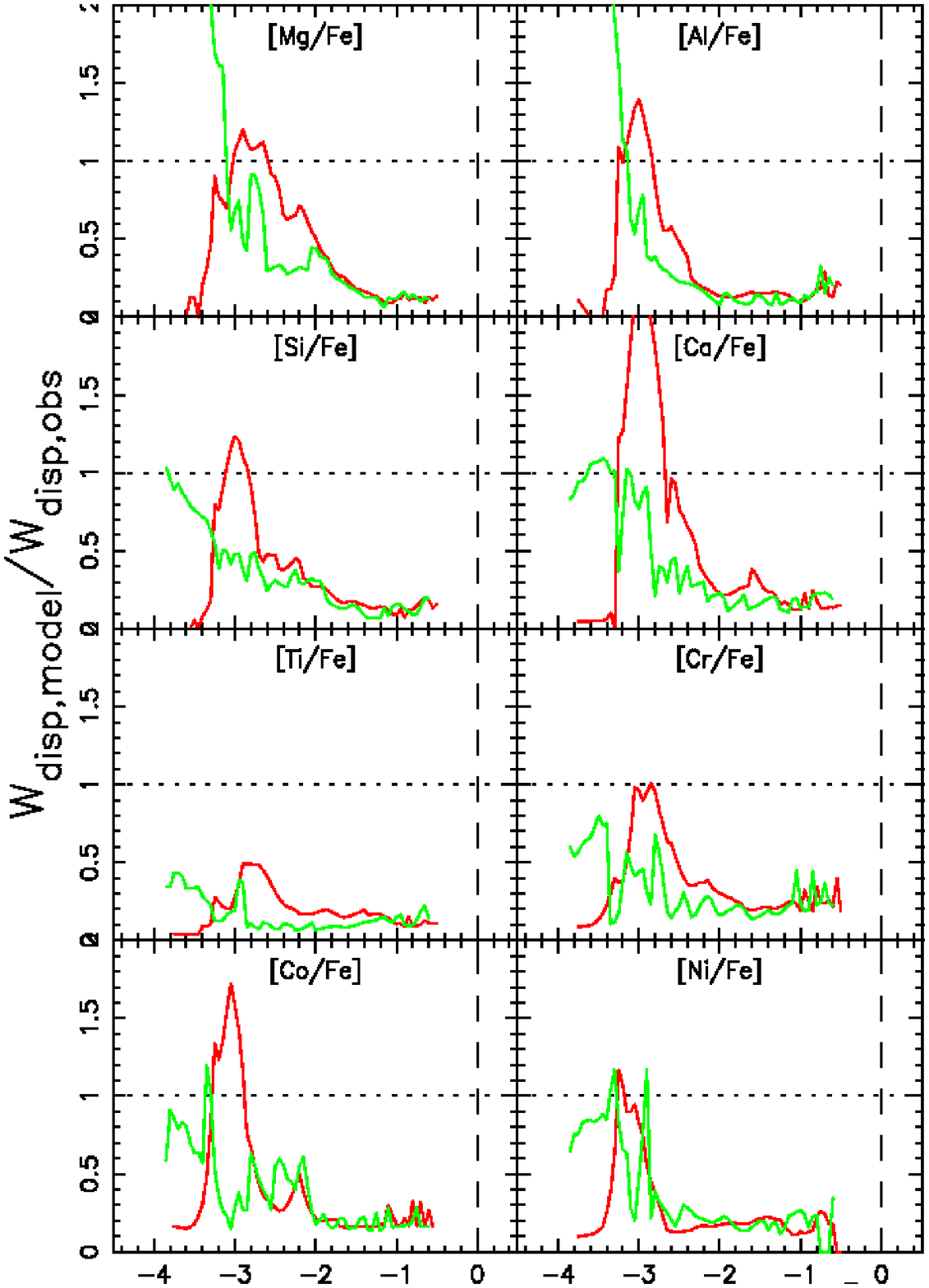,angle=0,height=7cm,width=\textwidth}
\caption{Dispersion (1 $\sigma$ around the mean) of abundance
ratios in the solar neighborhood.
Model results are obtained with 2 sets of yields: 
Nomoto et al. 1997 ({\it dark
curves}) and WW95 ({\it lighter curves}) and are divided by corresponding  
observed scatter as a function of metallicity (from Ishimaru et al. 2002).}
\label{fig:toosmall}
\end{minipage}
\end{figure}

In Fig. 1 (right panel)
 we compare our results of a detailed GCE model (Goswami and Prantzos 2000)
with observations from various sources, 
including the yet  unpublished data of Yong and Lambert (Yong, private 
communication).
The observational trends are, globally, reproduced by our model, although
the model isotopic ratios are systematically lower than observations.
This was also noticed in Timmes
et al. (1995). It may well be that the Woosley and Weaver (1995, WW95) 
yields underestimate
the importance of the neutron excess in the production
of the Mg isotopes or that the Yong and Lambert data systematically 
overestimate those ratios.  Another possibility is that
there is some other source of the neutron-rich Mg isotopes in the
late halo, like e.g. AGB stars with He-shells hot enough to
activate the $^{22}$Ne($\alpha$,n)$^{25}$Mg neutron source. This 
reaction would provide neutrons for the s-process in those
stars, but also produce large amounts of $^{25}$Mg and 
$^{26}$Mg. 


In any case, future observations of isotopic ratios at all metallicities
(like those of the Ti isotopes, Young and Lambert in preparation)
will bring a substantial refinement in our understanding of stellar
nucleosynthesis.

\section{(IM)PERFECT MIXING IN THE EARLY GALAXY?}

Standard models of GCE assume instantaneous and perfect mixing of the stellar
ejecta with the ISM; this leads to a unique value of metallicity (or of the
abundance ratio of any two elements) at all times. This simplistic assumption
may not hold, since simple arguments suggest that complete mixing may require
many millions of years; during that time a new star may be born with its
composition affected mostly by the ejecta of a closeby supernova. A dispersion
of abundance ratios, increasing at earlier times (and lower metallicities)
is, naively, expected in that case.

Those ideas are supported by observations of s- and r- elements (e.g. Truran
et al. 2002). However, for elements up to the Fe-peak Carretta et al. (2002) 
find no hints for intrinsic dispersion down to [Fe/H]$\sim$-3 (Fig. 2). 
This result
is intriguing since simple models, both 1-D (Tsujimoto and Shigeyama 1998;
Karlsson and Gustafsson 2001) and 2-D (Argast et al. 2000, 2002), find that
an important scatter should be obtained at such low metallicities, its
amplitude depending on adopted yields and model assumptions.

\begin{figure*}[!t]
\begin{center}
\psfig{figure=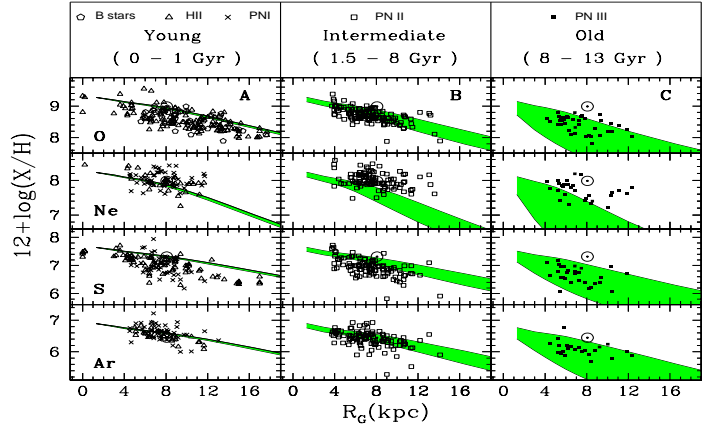,width=6cm,height=10cm,angle=-90}
\caption{Abundance gradients of O, Ne, S and Ar in planetary nebulae of
various age classes (from left to right) in MW disk; they are 
compared to model predictions of Hou et al. 2000 ({\it shaded aereas}).}
\end{center}
\end{figure*}

The results of recent 1-D models of inhomogeneous GCE (Ishimaru et al. 2002)
are displayed in Fig. 3. Two sets of stellar yields are used (Nomoto et al. 
1997 and WW95). The resulting dispersion is compared to the observational
one (estimated by Ryan et al. 1996). Depending on the adopted set of 
yields, a rather impotant dispersion is expected for $\alpha$-elements around 
[Fe/H]$\sim$-3 and below, while at metallicities higher than [Fe/H]=-2 
the theoretical scatter is extremely small. Note that uncertainties introduced by
observational errors ($\sim$0.1 dex) are not taken into account in the displayed model
results.

In the future, systematic surveys of low metallicity stars will help to
constrain stellar yields ({\it and} their dependence on stellar mass) and to
a better understanding of mixing processes in the ISM.

\section{ABUNDANCES IN THE MILKY WAY DISK}

Observed profiles of MW disk (gas, stars, SFR and abundances) 
require a larger SF efficiency and/or a smaller timescale of formation
(through infall) in the inner disk
(Boissier and Prantzos 1999). Models reproducing the above observational
constraints with few parameters predict that abundance gradients were
steeper in the past (e.g. Hou et al. 2000); they became progressively
flatter due to the inside-out formation of the disk (see Fig. 4).
That prediction is
in agreement with the recent observational results of Maciel and 
da Costa 2001 (see Fig. 5),
based on oxygen  abundances of planetary nebulae of various age classes 
[{\it Note}: taking into account the 
uncertainties in evaluating ages of planetary nebulae, the importance of that
agreement should not be overestimated; it is, however, encouraging].

\begin{figure}
\begin{minipage}[t]{75mm}
\psfig{figure=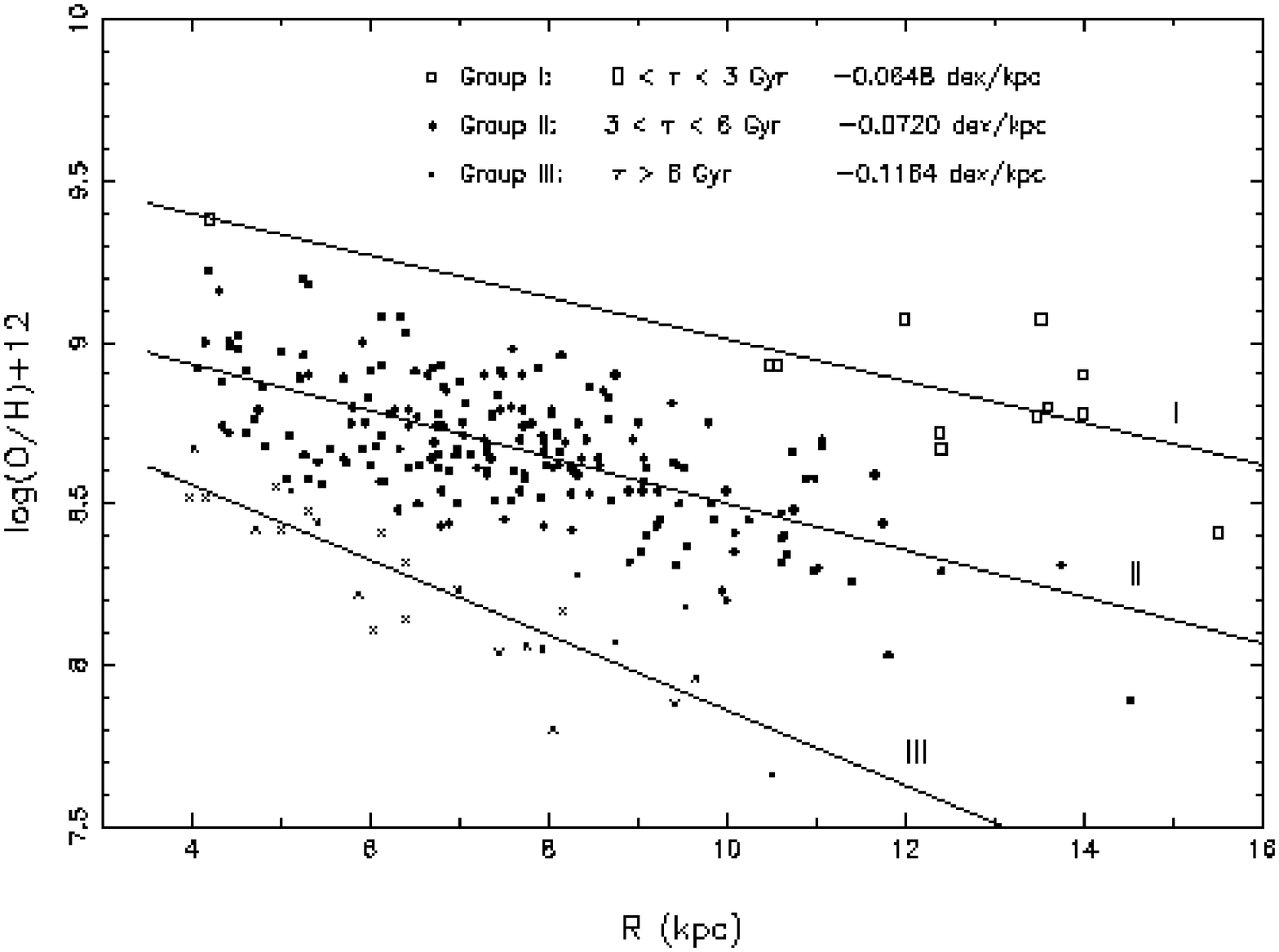,angle=90,height=6.cm,width=\textwidth}
\caption{Abundance gradients of O in MW disk, as traced by planetary nebulae
of various age  groups, according to Maciel and da Costa (2001).}
\label{fig:largenenough}
\end{minipage}
\hspace{\fill}
\begin{minipage}[t]{75mm}
\psfig{figure=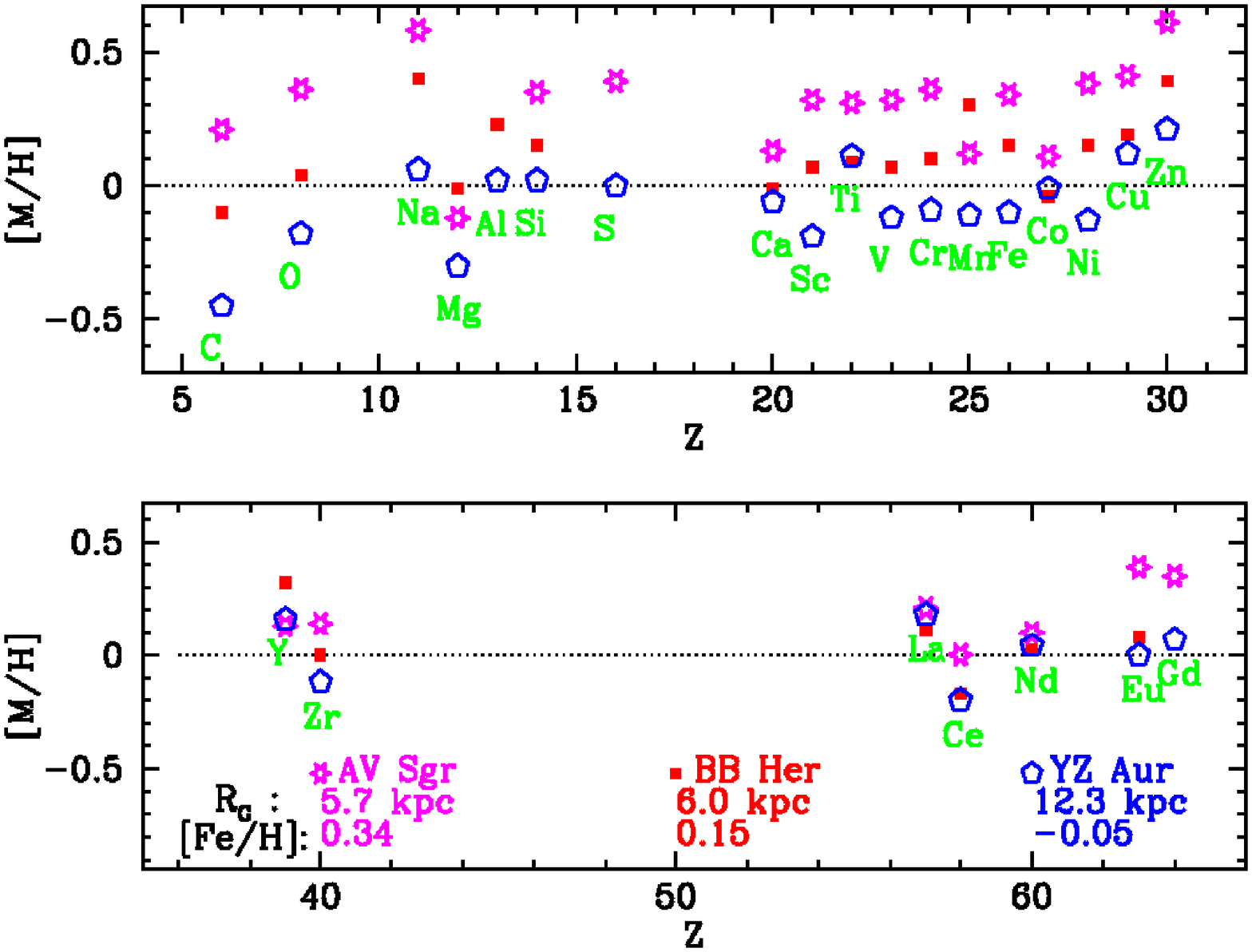,angle=90,height=6.cm,width=\textwidth}
\caption{Abundances (compared to solar) of 3 Cepheids in MW disk
(symbols in {\it lower panel}) from Andrievsky et al. (2002a).}
\label{fig:toosmall}
\end{minipage}
\end{figure}

It should be noted that the magnitude of the abundance gradient has been
questioned again, recently: Deharveng et al. (2000) and Pilyugin et al. (2002)
(based on abundances in HII regions) and  Cunha et al. 2002 (based on 
abundances of B-stars) find systematically smaller oxygen gradients than the
``canonical'' one of $\sim$-0.07 dex/kpc; clearly, the issue of the
abundance gradient in the Milky way (and other disk galaxies) is still
an open one.

In that respect, it should be noted that a recent study of abundances in 
Cepheids (with rather well determined distances) finds a bimodal slope
of the abundance profile, flatter in the solar neighborhood and steeper 
inwards (Andrievsky et al. 2002a,b). One of the important points of that work
is the determination of stellar abundance patterns concerning a large
number of elements (about 25). As can be seen in Fig. 6, there is a well defined
abundance pattern, scaled with metallicity, for elements up to the Fe peak
(upper panel), while no such pattern is evident for heavier elements (lower
panel). This is the first study establishing detailed abundance patterns in the
inner Milky Way; further works  of that kind in the future will
allow to study abundance patterns in those regions in the same way as we
study today detailed abundance patterns (mean trends, scatter, etc) in stars of
the galactic halo; interpretation of those patterns will require the 
calculation of detailed stellar yields up to metallicities of 3 Z$_{\odot}$.

\begin{figure}[t]
\begin{minipage}[t]{75mm}
\psfig{figure=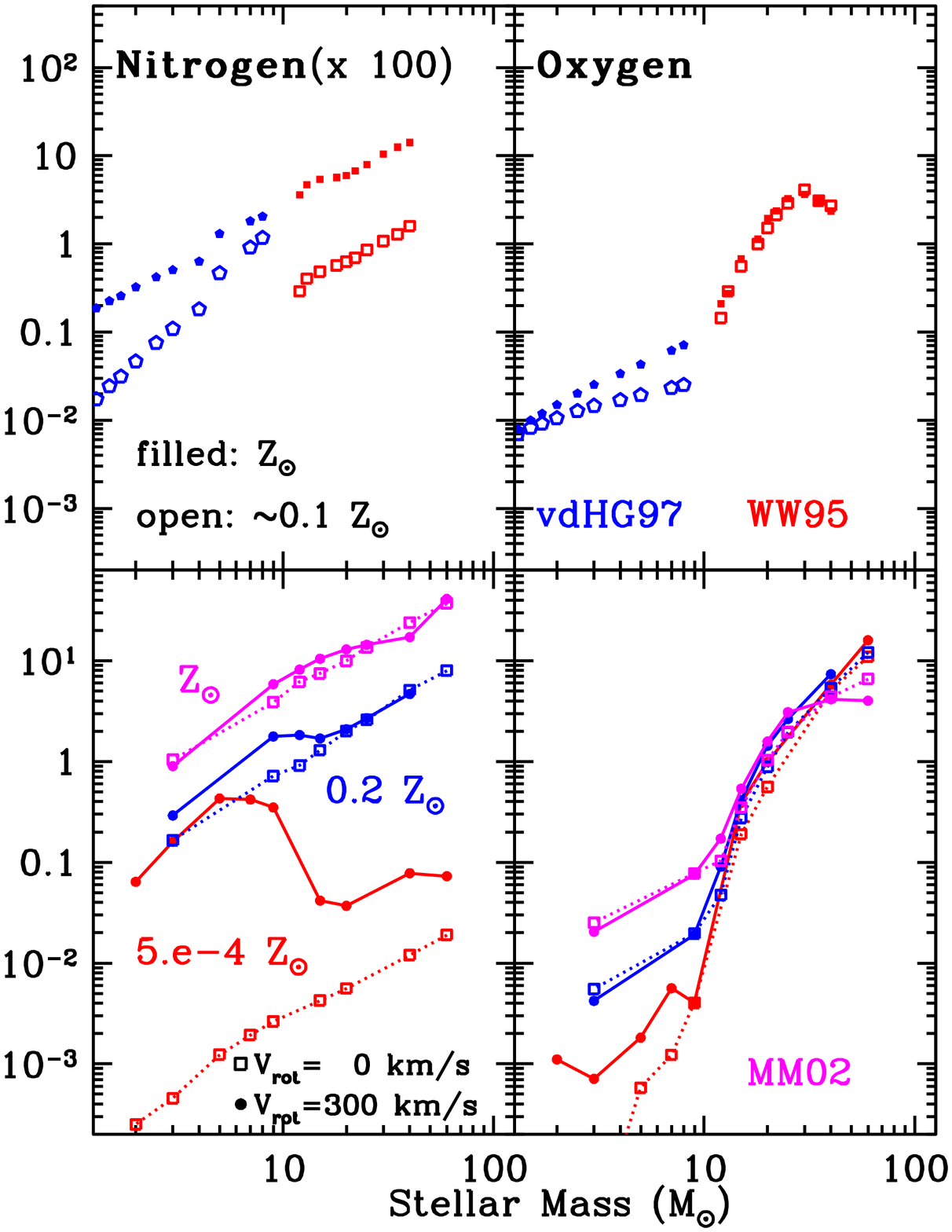,height=7cm,width=\textwidth}
\caption{Nitrogen and Oxygen yields. {\it Top}: Yields from van den Hoek
and Groenevegen 1997 (vdHG97, intermediate mass stars) and WW95 (massive stars).
{\it Bottom}: Yields from MM02 with rotation ({\it solid curves} and
no rotation ({\it dotted curves}). Yields (total ejected  masses) are
given for two (upper panel) and three (lower panel) initial metallicities.}
\label{fig:largenenough}
\end{minipage}
\hspace{\fill}
\begin{minipage}[t]{75mm}
\psfig{figure=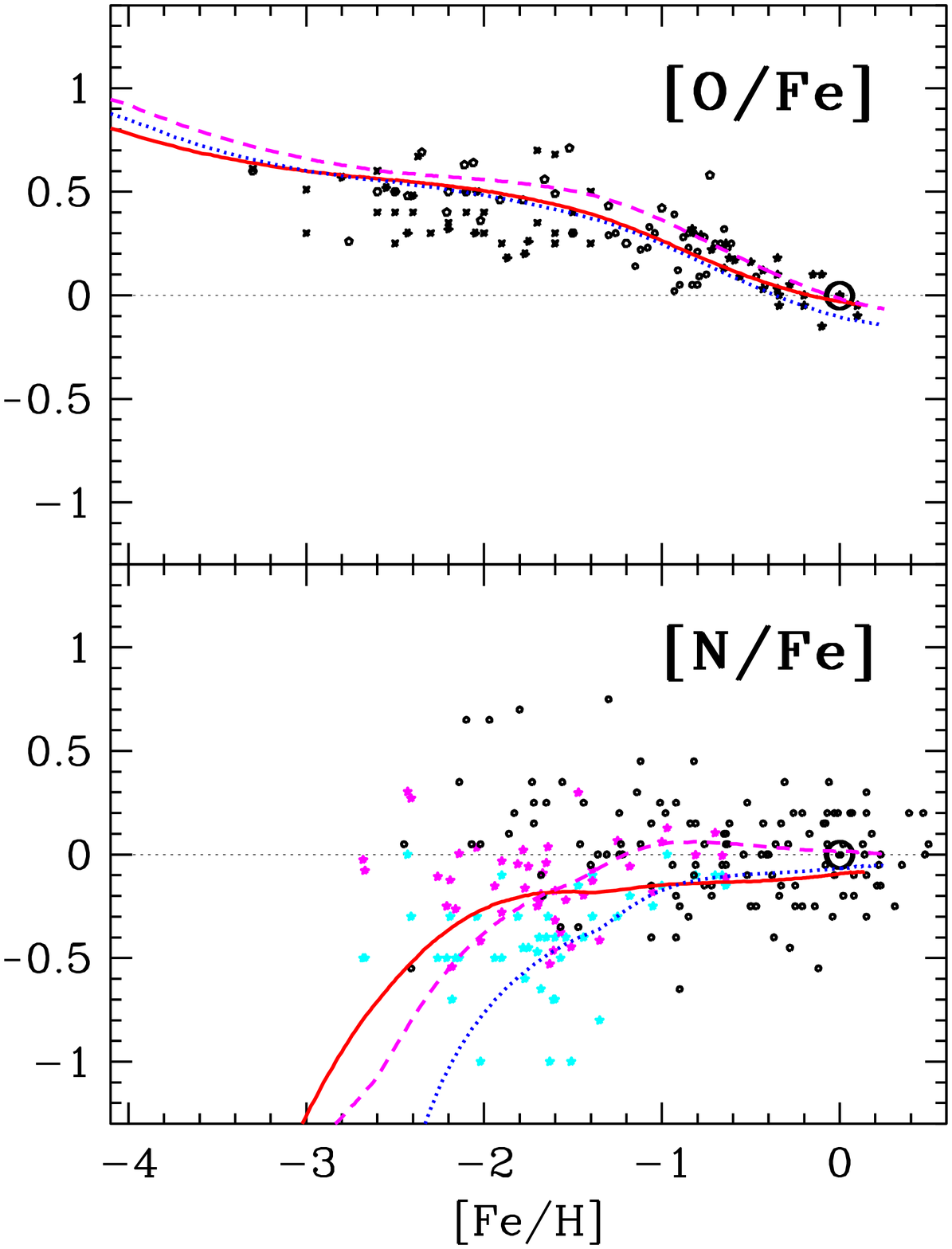,height=7cm,width=\textwidth}
\caption{Evolution of O and N in solar neighborhood with three sets of stellar
yields: vdHG97 + WW95 ({\it solid curve}), MM02 with rotation ({\it dashed
curves}) and MM02 with no rotation ({\it dotted curves}). The first two cases,
producing primary N from intermediate mass stars, lead to similar results.}
\label{fig:toosmall}
\end{minipage}
\end{figure}

\section{``PRIMARY'' NITROGEN IN THE EARLY GALAXY?}

Nitrogen is the main product of the CN cycle operating in core H-burning
in intermediate and high mass stars, in shell H-burning in stars of all masses
and in the putative Hot-Bottom Burning (HBB) in massive (4-8 M$_{\odot}$)
AGB stars. In the former two cases N is produced from the initial C (and O)
entering the stars, whereas in the latter from the C produced by the 3-$\alpha$
reaction in the He-shell (and brought to the envelope by convective motions
during thermal pulses). Thus, in the first two cases N is produced as a {\it
secondary} (its yield being quasi-proportional to initial stellar metallicity)
and in the latter as a {\it primary} (its yield being quasi-independent
of metallicity) [see top panel of Fig. 7].

No stellar evolution model finds HBB in a self-consistent way up to now: 
because of uncertainties in the treatment of convection the depth of the 
convective envelope has to be adjusted in order to reach the required
high temperatures (as e.g. in vdHG97). 
However, there is another way to produce primary N: 
rotationally induced mixing may mix protons to He-regions with abundant C
from the 3-$\alpha$ reaction. Self-consistent (1-D) models of rotating stars
have been recently calculated by Meynet and Maeder (2002, MM02). Their N
yields appear in Fig. 7 (bottom panel): at low metallicities,
substantial amounts of quasi-primary N are produced in intermediate mass 
rotating stars (with typical rotational velocities V$_{ROT}$=300 km/s).

The implications of those results for the evolution of N in the Galaxy are
shown in Fig. 8. The N yields of the rotating stars of MM02 lead to an 
evolution similar to the one obtained with the HBB yields of vdHG97: N
behaves as primary with respect to Fe, starting at [Fe/H]$\sim$--2.
With ``standard'' yields of non-rotating stars (with no HBB), N is
always produced as secondary, but it behaves
as primary w.r.t Fe (its  secondary production being matched by the delayed
Fe ejection from SNIa) only after [Fe/H]$\sim$--1.

Current obervations do not allow to conclude whether primary N is needed below
[Fe/H]$\sim$--2. If this were the case, primary N from massive stars 
(which dominate at such low metallicities and early times, due to their short
lifetimes) would be required; the corresponding mechanism has not
been found up to now.  Alternatively, the timescales of simple models
of GCE (like the one of Fig. 8), which are not seriously constrained by
other observations, should have to be revised.

\begin{figure}[t]
\begin{minipage}[t]{75mm}
\psfig{figure=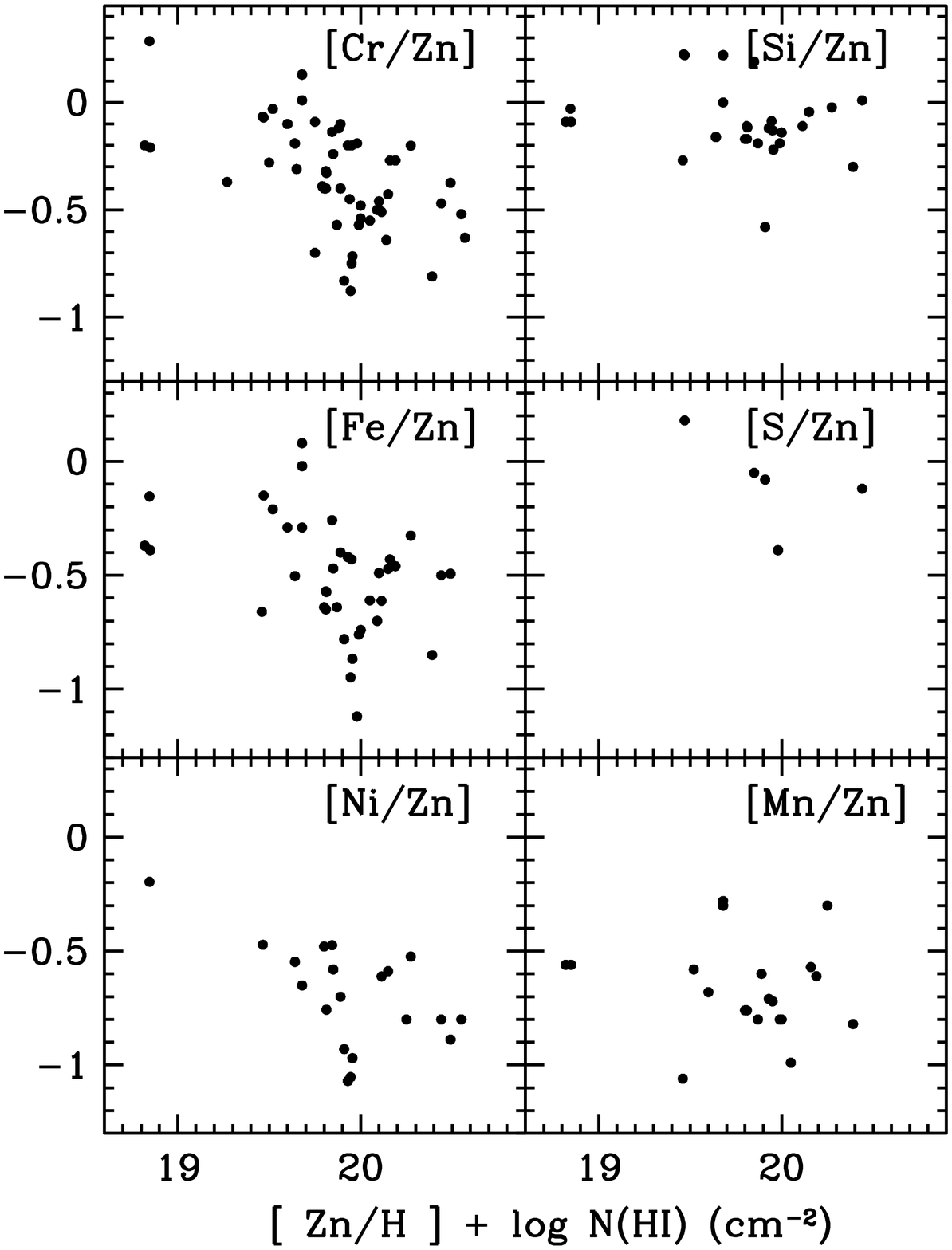,height=6cm,width=\textwidth}
\caption{Abundance ratios [M/Zn] in DLAs as a function of (a measure of) Zn
column density $\mathcal{F} = [Zn/H]+log(NHI)$. 
On the {\it left}: refractory elements have a declining
[M/Zn] with [Zn/H]+log(NHI); on the {\it right:} mildly refractory elements
show no such trend.}
\label{fig:largenenough}
\end{minipage}
\hspace{\fill}
\begin{minipage}[t]{75mm}
\psfig{figure=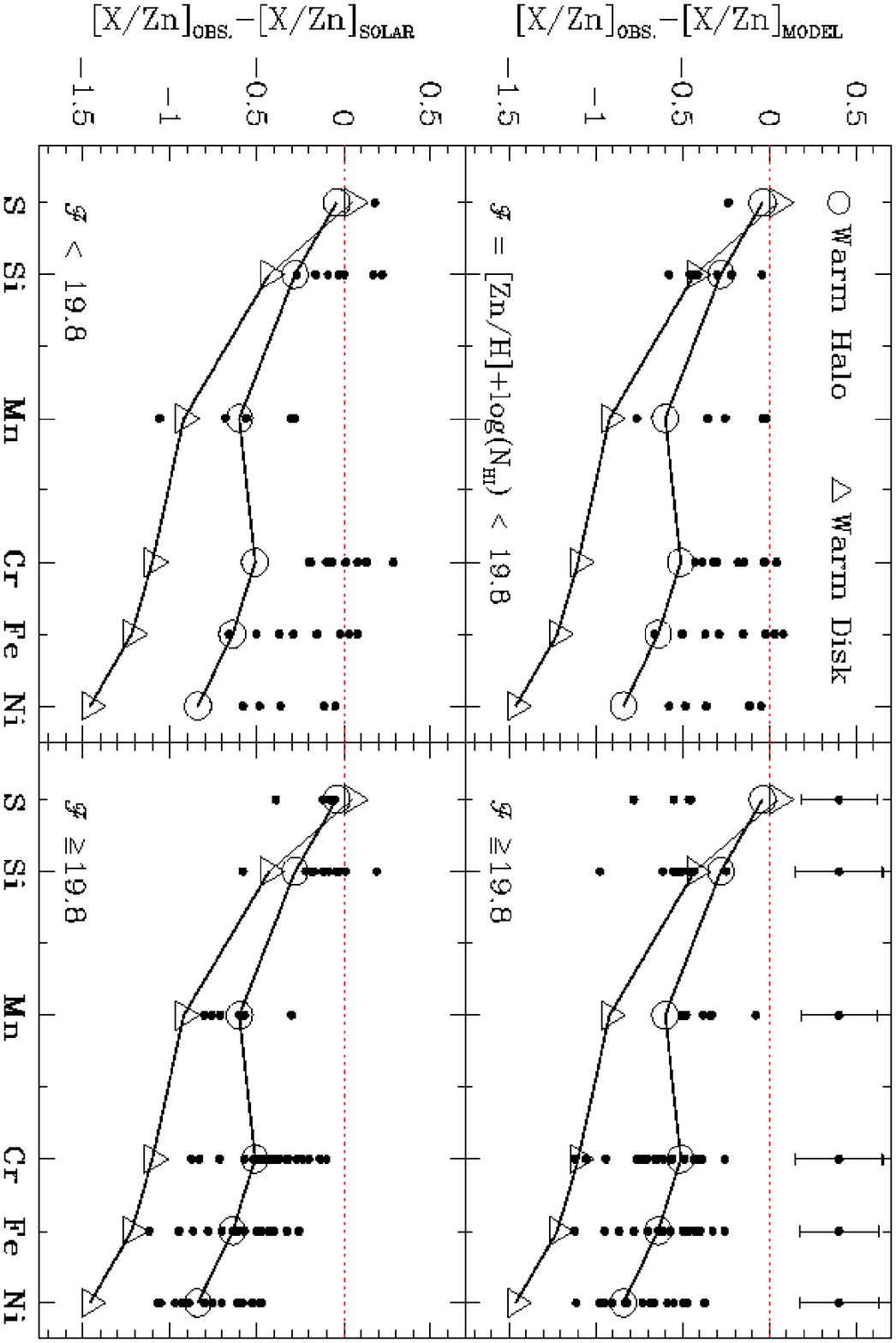,angle=90,height=6cm,width=\textwidth}
\caption{Depletion patterns in DLAs: abundance ratios [X/Zn] are compared 
to model results ({\it upper panels}) and to solar composition 
({\it lower panels}) for
low Zn column densities ({\it left}) and high ones({\it right}). The resulting
depletion pattern ressembles the one of warm MW halo clouds (from 
Hou et al. 2001).}
\label{fig:toosmall}
\end{minipage}
\end{figure}

\section{ABUNDANCE PATTERNS IN DLAs}

Damped Lyman $\alpha$ systems (DLAs) are high column density gaseous systems 
(N(HI)$>$2 10$^{20}$ cm$^{-2}$), detected through their absorption lines
in the optical spectra of quasars, up to relatively high redshifts
(up to $z\sim$5). Their study constitutes a powerful means to
investigate the properties of distant galaxies (or of their building
blocks). In particular, DLA metal abundances have been widely used
in the past few years to probe the nature of DLAs (e.g.  
Prochaska \& Wolfe 2002  and references therein).
However, it is not clear whether the observed abundances allow that,
because of various biases: depletion of metals into dust (Pei and Fall 1995)
or bias against too high or too low metal column densities (Boiss\'e 
et al. 1998). The latter bias, in particular, may explain the observed 
absence of evolution in the absolute abundance of Zn/H as a function
of redshift, as suggested in Prantzos and Boissier (2000).

Abundance ratios offer a better diagnostic tool than absolute 
abundances for the study of the chemical evolution of a system.
However, in the case of DLAs, the possibility of depletion into dust
grains (even to a small extent) makes difficult a direct interpretation
of the observed abundance patterns (see Vladilo 2002).

Hou et al. (2001)  found  an anticorrelation between
the observed abundance ratio X/Zn (where Zn is assumed to be undepleted
and X stands for the refractories  Fe, Cr and Ni) and metal
column density (Fig. 9). They suggested that this trend is an
unambiguous sign of depletion, since metal column density is
a measure of the amount of dust along  the line of sight.
Prochaska and Wolfe (2002) reached similar conclusions, finding that
the anticorrelation is also present if S is used instead of Zn.

Once dust depletion is unambiguously found, it is interesting to see
whether the depletion pattern ressembles to the ones encountered
in the various gaseous phases of the Milky Way (cold or warm clouds of the
halo or the disk, see Savage and Sembach 1996). 
Hou et al. (2001) found that :
if one assumes that the intrinsic metallicity pattern in DLAs is
the one given by their models (of disk galaxies), 
a difficulty arises with S, which is found to be
depleted at high $\mathcal{F} = [Zn/H]+log(NHI)$;
if, on the other hand, it is assumed that the
intrinsic DLA pattern is solar, 
a problem arises with Mn, which is found to be
depleted at low $\mathcal{F}$. In both cases the depletion patterns ressemble 
more to the one of warm halo clouds in the Milky Way (Fig. 10).

In any case, further studies of abundance patterns in DLAs will allow to probe
the nature of those systems and to study the early phases of cosmic
chemical evolution.

\end{document}